\documentclass[pre, amssymb, amsmath, nofootinbib, 10pt, twocolumn]{revtex4-1}

	\usepackage{graphicx}
	\usepackage{color}
	\usepackage{verbatim}
	\usepackage{subfigure}
	\usepackage{amsmath}
	\usepackage{amssymb}

\begin{document}
	
	\title{Manifest and Subtle Cyclic Behavior in Nonequilibrium Steady States }
	\author{R K P Zia$^{1,2}$, Jeffrey B Weiss$^3$, Dibyendu Mandal$^{3,4}$ and Baylor Fox-Kemper$^5$ }
	\affiliation{
	$^1$Department of Physics and Astronomy, Iowa State University, Ames, IA 50011, USA\\ 
	$^2$Department of Physics, Virginia Polytechnic Institute and State University, Blacksburg, VA 24061, USA\\	
	$^3$Department of Atmospheric and Oceanic Sciences, University of Colorado, Boulder, CO 80309, USA\\
	$^4$Department of Physics, University of California, Berkeley, CA 94720, USA\\
	$^5$Department of Earth, Environmental, and Planetary Sciences, Brown University, Providence, RI 02912, USA}
	
	\email{rkpzia@vt.edu}
	
	\begin{abstract}
	
Many interesting phenomena in nature are described by stochastic processes
with irreversible dynamics. To model these phenomena, we focus on a master
equation or a Fokker-Planck equation with rates which violate detailed
balance. When the system settles in a stationary state, it will be a
nonequilibrium steady state (NESS), with time independent probability
distribution as well as persistent probability current loops. The observable
consequences of the latter are explored. In particular, cyclic behavior of
some form must be present: some are prominent and manifest, while others are
more obscure and subtle. We present a theoretical framework to analyze such
properties, introducing the notion of \textquotedblleft probability angular
momentum\textquotedblright\ and its distribution. Using several examples, we
illustrate the manifest and subtle categories and how best to distinguish
between them. These techniques can be applied to reveal the NESS nature of a
wide range of systems in a large variety of areas. We illustrate with one
application: variability of ocean heat content in our climate system.
 	
	\end{abstract}
	
	\maketitle

\section{Introduction}

In the standard Boltzmann-Gibbs formulation of equilibrium statistical
mechanics, time plays no role. Once $\left\{ q\right\} $, the set of
configurations (or microstates) of the system of interest is chosen, the
energy functional (Hamiltonian) $H\left( q\right) $ is provided, and the
conditions for equilibrium are specified, then $P\left( q\right) $, the
probability for finding the system in $q$ is known. This structure is built
on Boltzmann's fundamental hypothesis: 
$P\left( q\right) \propto \delta \left(E-H\left( q\right) \right) $ 
for an isolated system with total energy $E$.
From here, various other ensembles and their associated $P$'s follow. The
main task is to compute averages of various observable quantities $\mathcal{O%
}\left( q\right) $, \thinspace $\left\langle \mathcal{O}\right\rangle \equiv
\Sigma _{q}\mathcal{O}\left( q\right) P\left( q\right) $. 

	The Boltzmann-Gibbs paradigm is clearly inadequate to describe many stochastic processes in nature.
	In addition to a need to describe time dependent phenomena (e.g., autocorrelations), there are many systems which interact with the environment in a manner that \textit{violates} time reversal. 
	In particular, \textit{all} biological systems consume nutrients and discard waste, with processes that clearly cannot be reversed in time. In this case, the system settles into \textit{non-equilibrium steady states }(NESS), with characteristics absent from systems in thermal equilibrium. 
	Chief among these are the presence of probability currents and loops \cite{ZS2007}, much like those in magnetostatics. 
	In this brief note, we report some observable consequences of these current loops, in both manifestly cyclic behavior and more subtle realizations. 
	We will also point to the notion of \textquotedblleft probability angular momentum\textquotedblright\ and relate it to a more familiar quantity: the two point correlation at \textit{unequal} times.

	For the simplest stochastic process that leads to NESS, we may start with a
master equation for the time dependent $P\left( q,t\right) $, with rates
that violate detailed balance (DB). 
	In general, this equation takes the form
	\begin{eqnarray}
	\label{eq:Master}
	\partial _{t}P\left( q,t\right) & = & \sum_{q^{\prime}}\left[ W\left(
q^{\prime }\rightarrow q\right) P\left( q^{\prime },t\right) -W\left(
q\rightarrow q^{\prime }\right) P\left( q,t\right) \right] \nonumber \\
	& \equiv & \sum_{q'} K(q' \rightarrow q),
	\end{eqnarray}
where $W\left(q^{\prime }\rightarrow q\right) $ is the transition rate for the system in $q^{\prime }$ to become one in $q$ and $K(q' \rightarrow q)$ is the \textit{net} probability current from $q^{\prime }$ to $q$. 
	Now, DB is often displayed as $W\left( q^{\prime }\rightarrow q\right) /W\left( q\rightarrow q^{\prime }\right) =\exp \beta \left[ H\left( q^{\prime }\right) -H\left(
q\right) \right]$, so that, for a system in thermal equilibrium with stationary $P^{\ast }\propto e^{-\beta H}$, the currents $K^{\ast }$ vanish identically. 
	(Note that quantities in the stationary state are associated with $^{\ast }$.)
	By contrast, for processes modeled by $W$'s that violate DB, some $K^{\ast }$'s in the NESS must be non-zero and must form closed loops.

	Though the master equation is the most general formulation for this class of stochastic processes, we will restrict ourselves, for simplicity, to configuration spaces described by continuous variables, $\xi _{\alpha }$ (or just $\vec{\xi}$) in arbitrary dimensions, and evolution controlled by the Fokker-Planck equation (FPE): \footnote{The Einstein summation convention is used here.} 
	\begin{equation}
	\label{eq:FPE}
	\partial _{t}P\left( \vec{\xi},t\right) =\partial ^{\alpha }\left\{ \partial ^{\beta }D_{\alpha \beta }P-V_{\alpha}P\right\} \equiv - \partial^\alpha K_\alpha.
	\end{equation}
	Here, $D$ and $V$ represent the diffusive and drift aspects, respectively. 
	A more intuitive description, as well as rules for coding simulations, is the Langevin equation 
	\begin{equation}
	\label{eq:Langevin}
	\partial _{t}\vec{\xi}=\vec{V}+\vec{\eta},
	\end{equation}
where $\vec{\eta}$ is a Gaussian noise with 	$\left\langle \vec{\eta}\right\rangle = 0$  and $\left\langle \eta_{\alpha}(t) \eta_{\beta}(t') \right\rangle = D_{\alpha \beta} \delta(t - t')$.
	As the FPE is just a continuity equation, we can identify the probability current here as $K_{\alpha }=-\left\{ \partial ^{\beta} D_{\alpha \beta }P - V_{\alpha }P\right\}$. 
	Of course, the stationary distribution, $P^{\ast }$, satisfies $\partial _{t}P^{\ast }=0$, i.e., $\vec{\nabla}\cdot \vec{K}^{\ast }=0$. 
	In this approach, the dynamics satisfies DB provided 
	$ \left[ D^{-1}\right] ^{\gamma \alpha }\left( \partial^{\beta }D_{\alpha \beta }-V_{\alpha }\right) $
	is the gradient of some scalar function $s\left( \vec{\xi}\right)$, i.e., $\partial^\gamma s$.
	Then, it is straightforward to show that $\vec{K}^{\ast }\equiv 0$ with $P^{\ast }\propto e^{-s}$. 
	Our interest here are processes which violate this condition, when $\vec{K}^{\ast }$ is non-trivial. 
	Being divergenceless, it can be expressed (in 3 dimensions) as curl of $\vec{\psi}^{\ast }$, the stream function (in the language of fluid dynamics), while $\vec{\nabla}\times \vec{K}^{\ast }$ is known as the vorticity, $\vec{\omega}^{\ast }$.

\section{Mass/fluid \textit{vs.} probability angular momenta; two point
correlation functions}

	Angular momentum, a familiar concept from textbooks on classical mechanics and fluid dynamics, is associated with mass in motion: $\vec{L}=\vec{r} \times m\vec{v}$, $\int \mathrm{d} \vec{r} \vec{r}\times \vec{v}\rho \left( \vec{r}\right) $, and $\int \mathrm{d} \vec{r} \left( \vec{r}\times \vec{J} \right)$ (where $\vec{J}=\rho \vec{v}$ is the fluid current). 
	We transfer this fluids concept to probability by considering the mapping $\{ \vec{r}, \rho, \vec{J} \} \rightarrow \{ \vec{\xi}, P, \vec{K} \}$, so that $\int \mathrm{d} \vec{\xi} \, \left(\vec{\xi}\times \vec{K} \right)$ is a quantity of interest and naturally named \textquotedblleft probability angular momentum.\textquotedblright\ 
	Of course, in arbitrary dimensions, rotations and angular momenta are not
vectors, but pseudo tensors. 
	Thus, instead of a vector $\vec{L}$, we should consider a matrix $\mathbb{L}$, with elements $L_{\alpha \beta }=-L_{\beta \alpha }$ and  
	\begin{equation}
	L_{\alpha \beta }=\int \mathrm{d}\vec{\xi} \, \left( \xi _{\alpha }K_{\beta } - \xi _{\beta}K_{\alpha } \right).  
	\label{Lab}
	\end{equation}
	Note that, due to the normalization condition $\int \mathrm{d} \vec{\xi} \,   P=1$, our \textquotedblleft mass\textquotedblright\ is unity, so that the unit of $\mathbb{L}$ is just $\xi ^{2}/t$, precisely that of diffusion. 
	This remarkable feature is not coincidental, as the intimate connection between them will be shown in the next section. 
	We further note that, in a NESS, we have $\int \mathrm{d} \vec{\xi} \,  \vec{K}^{\ast }=0$, so that $\mathbb{L}^{\ast }$ is independent of the choice of the origin of $\vec{\xi}$ (and is just the total vorticity). 
	Other familiar concepts such as angular velocity $\vec{\Omega}$ and inertia tensor, $\mathbb{I}$, also have analogs. 
	Specifically, we see that $\int r_{i}r_{j}\rho $ maps to the two point correlation $\left\langle \xi _{\alpha }\xi _{\beta }\right\rangle \equiv \int \xi_{\alpha }\xi _{\beta }P$. 
	Meanwhile, in analog with $\vec{v}=\vec{\Omega}\times \vec{r}$, we can define our angular velocity via $K_{\alpha }=\Omega_{\alpha }^{\beta }\xi _{\beta }P$ \cite{JBW2007} and come up with the equivalent of $\vec{L}=\mathbb{I}\vec{\Omega}$.

	We end this section with a not-so-familiar generalization of angular momentum. 
	If the trajectory of a point mass $\vec{r}\left(t\right)$ is known, then we may consider the quantity $\vec{a}\equiv \vec{r}\left( t\right)\times \vec{r}\left( t^{\prime }\right) $, the magnitude of which is just the area of the parallelogram spanned by the two vectors. 
	Clearly, it is related to the angular momentum by $\vec{L}=\left. m\partial _{t^{\prime }}\vec{a}\right\vert _{t^{\prime }=t}$. 
	The analogous generalization to $L_{\alpha \beta }$ here is a special two point correlation function: 
	\begin{equation}
	\label{eq:CTensor}
	\tilde{C}_{\alpha \beta }\left( t,t^{\prime }\right)\equiv \int \mathrm{d} \vec{\xi} \, \int \mathrm{d} \vec{\xi'} \, \mathcal{P}\left( \vec{\xi}, t; \vec{\xi^{\prime}}, t^{\prime } \right) \left[ \xi _{\alpha }\xi_{\beta }^{\prime } - \xi _{\beta } \xi_{\alpha }^{\prime }\right],
	\end{equation}
where $\mathcal{P}$ is the \textit{joint} probability. 
	If $t^{\prime }>t$, then $\mathcal{P}$ is the product of the conditional
probability, $\mathcal{G}\left( \vec{\xi ^{\prime }},t^{\prime }| \vec{\xi}, t\right)$, and $P\left( \vec{\xi}, t \right)$. 
	Regarding the FPE (Eq.~\ref{eq:FPE}) as a Schr\"{o}dinger equation, $\mathcal{G}$ is the familiar propagator in quantum mechanics. 
	As in the point mass case, we have 
	\begin{equation}
	\label{eq:LC}
	L_{\alpha \beta }=\left. \partial _{t^{\prime}}\tilde{C}_{\alpha \beta }\right\vert _{t^{\prime }=t}.
	\end{equation} 
	Meanwhile, we see that $\tilde{C}_{\alpha \beta }\left( t,t^{\prime }\right) $ is just twice the antisymmetric part of the two point correlation at arbitrary times, $\left\langle \xi _{\alpha }\xi _{\beta }\right\rangle _{tt^{\prime }}$.
	Being antisymmetric, it is necessarily odd under exchange $t\Leftrightarrow t^{\prime }$. 
	In the stationary state, translation invariance prevails and so, $\tilde{\mathbb{C}}^{\ast }$ depends only on the difference $\tau \equiv t^{\prime }-t$. 
	Since it is \textit{odd }under time reversal, $\tilde{\mathbb{C}}^{\ast}\neq 0$ is a concrete measure of DB violation and irreversibility in a stationary state.

	While the framework presented above is valid for all stochastic processes, it is valuable to illustrate these ideas in an explicitly solvable system: the Linear Gaussian Model (LGM) \cite{Lax1966,JBW2007,ZS2007}.
	In an LGM, $D_{\alpha \beta }$ (elements of $\mathbb{D}$) are constants, while $\vec{V}$ is linear in $\vec{\xi}$, as in generalized simple harmonic oscillators (SHO): $\vec{V}=\mathbb{A}\vec{\xi}$. 
	(Refer to Eqs.~\ref{eq:FPE} and \ref{eq:Langevin}.)
	Thus, the model is completely defined by two matrices, $\mathbb{D}$ and $\mathbb{A}$. 
	Of course, $\mathbb{D}$ must be positive symmetric, while the real parts of the eigenvalues of $\mathbb{A}$ must be negative (for the stability of the process). 
	If $\mathbb{D}^{-1}\mathbb{A}$ is symmetric, then DB is satisfied and the scalar $s$  is $-\vec{\xi}\cdot \mathbb{D}^{-1}\mathbb{A}\vec{\xi}/2$. 
	If not, then the stationary $P^{\ast }$ is still a Gaussian\cite{Lax1966}: $P^* \propto \exp \left\{ -\vec{\xi}\cdot \mathbb{C}^{-1}\vec{\xi}/2\right\} $,
where $\mathbb{C}$ is fixed by \cite{JBW2007} $\mathcal{S}\left[ \mathbb{AC}\right] =-\mathbb{D}$ and $\mathcal{S}$ stands for \textquotedblleft the symmetric part of.\textquotedblright\ 
	Clearly, $\mathbb{C}$ is the covariance matrix in the steady state, i.e., the equal time two point function. (Note that $\mathbb{C}$ is not the same as $\tilde{\mathbb{C}}$!)
	Meanwhile, we have $\vec{\xi}P^{\ast } = -\mathbb{C}\vec{\nabla}P^{\ast }$, which leads to $\vec{K}^{\ast} = - \left\{ \mathbb{D} \vec{\nabla} - \mathbb{A} \vec{\xi} \right\} P^{\ast} = - \left[ \mathbb{AC + D} \right] \vec{\nabla}P^{\ast}$.
	Since $\mathbb{D}=-\mathcal{S}\left[ \mathbb{AC}\right] $, the sum $\left[ \mathbb{AC + D} \right] $ is $\mathcal{A}\left[ \mathbb{AC}\right] $, the \textit{antisymmetric} part of $\mathbb{AC}$. 
	Thus, $\vec{K}^{\ast }$ is manifestly divergence free while the stream function (a matrix here) can be identified as $ -\left[ \mathbb{AC+D} \right] P^{\ast}$. 
	Moreover, it is straightforward to obtain an explicit expression for $\mathbb{L}^{\ast }$:
	\begin{equation}
	\mathbb{L}^{\ast }=-2\mathcal{A}\left[ \mathbb{AC}\right]   \label{L*}
	\end{equation}%
	and its generalization: 
	\begin{equation}
	\mathbb{\tilde{C}}^{\ast }\left( \tau \right) =-2%
	\mathcal{A}\left[ e^{\mathbb{A}\tau }\mathbb{C}\right]
	\end{equation}%
In this setting,
we see that diffusion ($\mathbb{D}$) and angular momentum 
($\mathbb{L}^{\ast }/2$)
are just the symmetric and antisymmetric parts of one matrix: $-\mathbb{AC}$. 
	Thus, they must have the same units, and play complementary roles in any stochastic process. 
	Finally, note that the angular velocity matrix is given by $\Omega ^{\ast } = \mathbb{DC}^{-1}+\mathbb{A}$ \cite{JBW2007}, while $\mathbb{L}^{\ast } = 2\mathbb{C}\Omega ^{\ast }$ is the analog of $\vec{L} = \mathbb{I}\vec{\Omega}$.

\section{Distributions of $\mathbb{L}$}

	Having established that there must be some cyclic behavior in any NESS, we ask if this feature is displayed (a) prominently and manifestly, or (b) in some obscure and subtle way. 
	To answer which category a system belongs to, we must go beyond the average values of angular momenta (Eqs.~\ref{Lab} and \ref{L*}) and study the full distribution: $p\left( L_{\alpha \beta }\right) \equiv \int \delta \left( L_{\alpha \beta }-\left[ \xi_{\alpha }K_{\beta} - \xi _{\beta }K_{\alpha }\right] \right)$. 
	For simplicity, let us consider a two-dimensional $\vec{\xi}$-space, so that there is only one independent component in any antisymmetric matrix, e.g., $\left\langle \mathcal{L}\right\rangle \equiv L_{12}^{\ast }$. 
	Then, our task simplifies to the study of $p\left( \mathcal{L}\right)$. 
	From simulations, it can be obtained by computing $\mathcal{L}\left( t\right) $ from a long trajectory $\vec{\xi} \left( t\right) $ (in the steady state), and compiling a histogram.

	\begin{figure}
	\begin{center}
	\includegraphics[width = 7 cm]{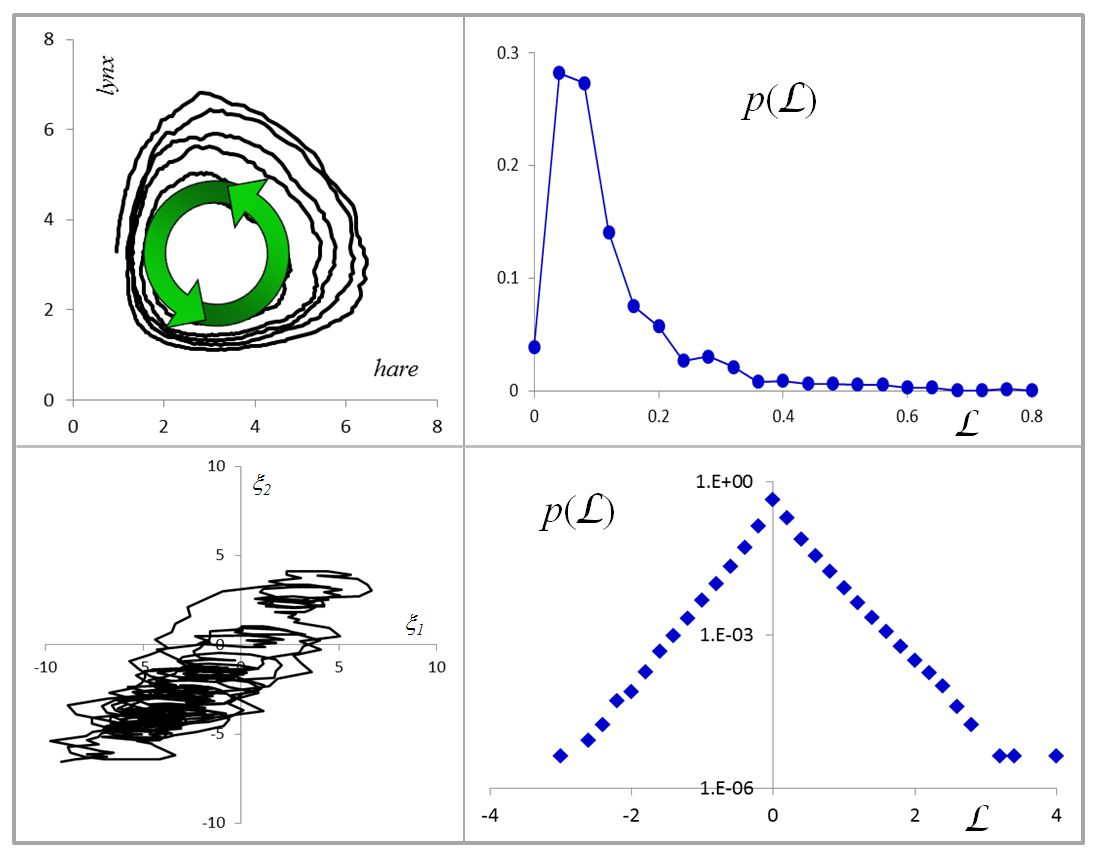}
	\end{center}
	\caption{\label{Fig1} Short trajectories and distributions $p\left( \mathcal{L} \right)$ for two simple stochastic processes: noisy Lotka-Volterra (upper panels) and SHO's coupled to thermal baths at different temperatures (lower panels). In the former, cyclic behavior is manifest, associated with a \textquotedblleft one-sided\textquotedblright  $p\left( \mathcal{L} \right)$. By contrast, this behavior is quite subtle in the latter system, with a distribution that is almost symmetric (around $\mathcal{L}=0$).}
	\end{figure}

	Clearly, the most extreme example in category (a) is a deterministic orbit (e.g., Keplerian) which yields a fixed angular momentum: $p=\delta \left( \mathcal{L} - const \right)$. 
	More common, stochastic systems of this type will display broader $p$'s. 
	To illustrate, we consider a stochastic Lotka-Volterra model \cite{LV2026} for the population of hares and lynx, $\xi _{h}$ and $\xi_l$ respectively. A specific example,
\footnote{We caution that this model is designed to illustrate properties in category (a) systems. It is too simplistic to provide a good description of the full complexity of predator-prey behavior. See \cite{MN2005} for a good treatment. For a recent review on stochastic LV systems, see, e.g., \cite{MGT2007}.}
	associated with the upper panels of Fig. 1, is:
\begin{equation}
\dot{\xi}_{h}=\xi _{h}\left[ 2-\xi _{l}\right] ,~~~\dot{\xi}_{l}=\xi _{l} \left[ -4+\xi _{h}\right] 
\end{equation}
plus noise. 
	The figure shows a typical trajectory in the space of hare/lynx populations from a simple simulation run, as well as the associated distribution $p\left( \mathcal{L} \right)$. 
 	Clearly, the latter is the result of noise on a $\delta $ distribution. 
	Since the dynamics of such models manifestly violate time reversal symmetry and DB, a distribution dominated by one sign of $\mathcal{L}$ is expected. 
	Turning to the subtler NESS in category (b), the average $\left\langle \mathcal{L} \right\rangle$ can be quite small, the result of a broad $p\left( \mathcal{L}\right)$ with only a slight asymmetry in $\mathcal{L}$. 
	Let us illustrate with two SHO's coupled to thermal baths at different $T$'s \cite{ZT1998}.
	Starting with the standard Hamiltonian $\mathcal{H}=\left[ k_{1}\xi_{1}^{2}+k_{2}\xi _{2}^{2}+k_{\times }\left( \xi _{1}-\xi _{2}\right) ^{2}\right] /2,$ we model the effects of the thermal baths by 
	\begin{equation}
	\label{eq:TTSHO}
	\dot{\xi}_{\alpha} = -\lambda _{\alpha }\left( \partial \mathcal{H}/\partial \xi _{\alpha}\right) +\eta _{\alpha }, 
	\end{equation}
with $\left\langle \eta \right\rangle =0$ and $\left\langle \eta _{\alpha}(t) \eta _{\beta}(t') \right\rangle = 2 \lambda_{\alpha} T_{\alpha} \delta_{\alpha \beta} \delta(t - t')$. 			Since this system is precisely an LGM, we find $\left\langle \mathcal{L}\right\rangle \propto k_{\times }\left( T_{1}-T_{2}\right) $, which shows that it settles into a NESS only when $k_{\times }\neq 0$ and $T_{1}\neq T_{2}$. 
	In the lower left panel of Fig.~1, we show a typical trajectory, which displays no obvious preferential rotation. 
	The right panel shows a nearly symmetric $p\left( \mathcal{L}\right)$, with barely discernible asymmetry. 
	As a result, the average $\left\langle \mathcal{L}\right\rangle \thicksim 4.70\times 10^{-3}$ is much smaller than the standard deviation $\Delta \mathcal{L}\thicksim 32.41 \times 10^{-3}$. 
	Here, we find $\left\langle \mathcal{L}\right\rangle $ to be entirely consistent with the theoretical prediction of $4.66\times 10^{-3}$.
	Finally, we note that the large $\mathcal{L}$ behaviors are exponential, governed by different decay constants. 
	These can also be computed, since we can obtain analytically the full $p \left( \mathcal{L} \right)$ of Gaussian models (details to be published elsewhere \cite{NMFWZ2016}).

\begin{figure}
\begin{center}
\includegraphics[width = 7 cm]{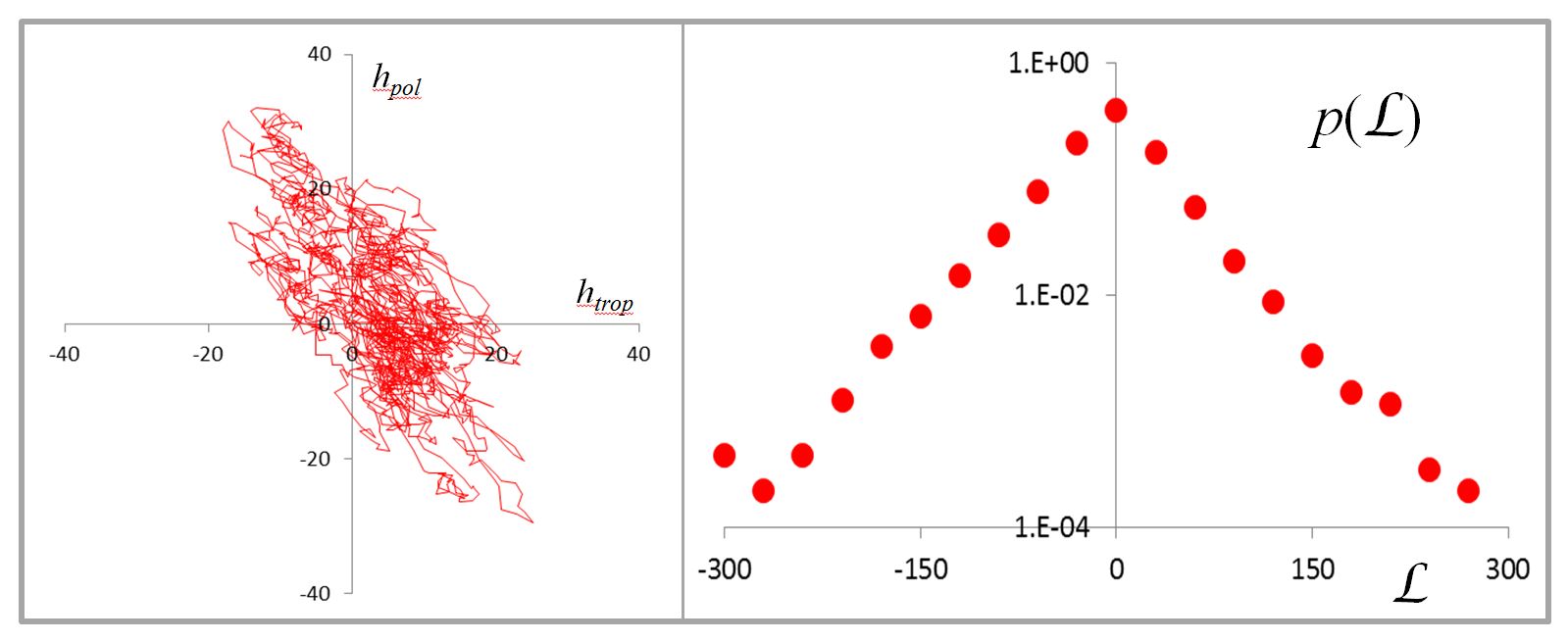}
\end{center}
\caption{\label{Fig2}
A short trajectory and distribution $p\left( \mathcal{L} \right)$ for 
the ocean heat content anomalies found in the tropical and polar regions 
(in a millennium long run with the Community Earth System Model). The units
of $h$ is $ZJ = 10^{21}$ Joules. The units for $\mathcal{L}$ is $ZJ^2$ per
season. The distribution indicates that cyclic behavior in these two
variables is very subtle.
}
\end{figure}

	Apart from these simple examples, we have recently studied other systems in NESS, including an epidemic model with asymmetric infection rates \cite{SZ2014} and a heterogeneous non-linear q-voter model \cite{MMZ2016}. 
	Both settle into NESS that display only very subtle cyclic behavior.
	Here, let us present preliminary results concerning a much larger and complex
system: variations in the heat content of our oceans. 
	As a stochastic process, the oceans are heated in the tropical regions and suffer loss
mostly from the polar regions, forming clearly a non-equilibrium system.
	Over long periods, it appears quasi-stationary and may be regarded as a NESS. 
	Unfortunately, high quality data for these anomalies in the real oceans form only a small set and date from about half a century ago~\cite{NOAA}.
	Nevertheless, we can combine the data into two time series, $h_\text{tropics}\left( t \right)$ and $h_\text{polar}\left( t\right) $, and study both $\left\langle \mathcal{L}\right\rangle$ and $p\left( \mathcal{L}\right)$.
	The results are very similar to those of the SHO's in Fig.~1. 
	The details are quite complicated and will be published elsewhere~\cite{NMFWZ2016}. 
	Here, we turn to a much longer (about a millennium) data set, created using the state-of-the-art Community Earth System Model~\cite{CESM}.
	Though non-trivial complications concerning
the analysis also exist here \cite{NMFWZ2016}, the results are consistent
with the those from real data. Illustrating with a small portion of this
trajectory and showing the $p\left( \mathcal{L}\right) $ in Fig. 2, we
recognize that these NESS aspects are similar to those in the
two-temperature SHO case. However, there is a subtle difference: 
$\left\langle \mathcal{L}\right\rangle \thicksim -5.4~ZJ^{2}/season$ is
negative, a sign naively opposite to that in the SHO's. This difference
indicates that temperature difference is not the only controlling factor 
for the sign of $\left\langle \mathcal{L}\right\rangle $. The more 
significant factor is the level of the noise associated with each 
degree of freedom. Details of this kind of analysis and understanding 
will be presented elsewhere \cite{NMFWZ2016}.

\section{Summary and Outlook}

We have shown that the angular momenta $\mathbb{L}^{\ast }$ and the two
point correlation at unequal times $\mathbb{\tilde{C}}^{\ast }\left( \tau
\right) $ are, for \textit{any} NESS, excellent measures of the underlying
time-reversal and DB violating dynamics. Beyond these simplest
quantities, any multipoint correlation functions at unequal times will
provide a platform for measuring such characteristics. 
In recent years, there has been other attempts to characterize cyclic 
behavior in NESS, e.g., those by Russell and Blythe \cite{RB2013}. These
approaches rely on the properties of angular displacements (in 
conffiguration space) rather than angular momenta. While there are some
advantages to the former (e.g., intuitively understandable, independence
of the scales of $\vec{\xi}$), there are also disadvantages 
(e.g., over-emphasis of motion near the origin, non-anayticity at 
$\vec{\xi}=0$). By contrast, $\mathbb{L}$ is analytically tractable, 
while it also enjoys intimate relationships with other key quantities, 
such as the probability current $\vec{K}$, the two point function 
$\mathbb{\tilde{C}}$, and the diffusion matrix $\mathbb{D}$. 

We have also shown that, for certain systems in NESS, DB
violation is so patently obvious that the irreversible nature is prominent
and manifest. For example, hares do \textit{not} prey on lynxes! 
On the other hand, many
NESS do not display such clear behavior, as the time-reversal violating
aspects are more obscure and subtle. Clearly, it is desirable to understand
more deeply the mechanisms which control the outcomes of a system. What are
the parameters which, when varied continuously, will take a system from
category (a) to (b)? Is the \textquotedblleft transition\textquotedblright\
abrupt and discontinuous? or smooth and continuous? Is it possible that 
$p\left( \mathcal{L}\right) $ displays a combination of both
\textquotedblleft components\textquotedblright ? We believe this is a
promising and rich avenue for future research, both as a novel measure to
characterize different systems in NESS and as a possible step towards a
overarching framework for the foundations of non-equilibrium statistical
mechanics.

\acknowledgements

We are grateful to A Mellor, M Mobilia, A D Nelson, B Schmittmann and U C 
T\"{a}uber for enlightening discussions. This research is supported in part by
US National Science Foundation grants: OCE-1245944 and DMR-1507371.


\end{document}